    \documentclass[a4paper]{IEEEtran}
	 \usepackage{graphicx}

	 \usepackage{amsmath}
	 \usepackage{amssymb}
	 \usepackage{amsfonts}
	 \usepackage{epsfig}

	 \usepackage{pdfpages}
	 \usepackage{amsmath}    
	 \usepackage{graphicx}   
	 \usepackage{color}      
	 \usepackage{scrextend}
     \usepackage[bookmarks=false]{hyperref}
	 \usepackage{comment}
	 \usepackage{subfigure}
	 \usepackage{times}
	 \usepackage{hyphenat}
	 \usepackage{epsfig}
	 \usepackage{latexsym}
	 \usepackage{bbold,bbm}
	 \usepackage{setspace}
	 \usepackage{amssymb,amsmath}
	 \usepackage{mathrsfs}
	 \usepackage{amsgen,amsfonts,amsbsy,amsthm}
	 \usepackage{algorithmic,algorithm}
	 \usepackage{array}
	 \usepackage{booktabs}
	 \usepackage{amsfonts}
	 \usepackage{amssymb}
	 \usepackage{multirow}
	 \usepackage{longtable}
	 \usepackage{epstopdf}
	 \usepackage{wrapfig}
	 \usepackage[font=scriptsize]{caption}
	 \usepackage{enumerate}
	 \usepackage{footmisc}
	 \usepackage{textcomp}
	 \usepackage[latin1]{inputenc}
\usepackage{tikz}
\usetikzlibrary{shapes,arrows}

	 \newtheorem{defn}{Definition}
	 \newtheorem{thm}{Theorem}

	 \newtheorem{cor}[thm]{Corollary}

	 \newcommand{\bit}{\begin{itemize}}
	 	\newcommand{\eit}{\end{itemize}}
	 \newcommand{\bcor}{\begin{cor}}
	 	\newcommand{\ecor}{\end{cor}}
	 \newcommand{\beq}{\begin{equation}}
	 	\newcommand{\eeq}{\end{equation}}
	 \newcommand{\beqn}{\begin{equation*}}
	 	\newcommand{\eeqn}{\end{equation*}}
	 \newcommand{\bea}{\begin{eqnarray}}
	 	\newcommand{\eea}{\end{eqnarray}}
	 \newcommand{\bean}{\begin{eqnarray*}}
	 	\newcommand{\eean}{\end{eqnarray*}}
	 \newcommand{\ben}{\begin{enumerate}}
	 	\newcommand{\een}{\end{enumerate}}
	 \newcommand{\bdefn}{\begin{defn}}
	 	
	 	\renewcommand\footnotemark{}

		 	\newcommand{\uh}{\underline{h}}

		\newcommand{\calc}{\ensuremath{\mathcal{C}}}
		\newcommand{\calcp}{\ensuremath{\mathcal{C}^{\perp}}}

\begin{document}
	\begin{NoHyper}
\pagestyle{empty}
	\sloppy
	\title{A Rate-Optimal Construction of Codes with Sequential Recovery with Low Block Length} 
%
	\author{
		\IEEEauthorblockN{Balaji Srinivasan Babu, Ganesh R.Kini and P. Vijay Kumar, \it{Fellow}, \it{IEEE}}
		
		\IEEEauthorblockA{Department of Electrical Communication Engineering, Indian Institute of Science, Bangalore.  \\ Email: \{balaji.profess, kiniganesh94,  pvk1729\}@gmail.com} 
		
		\thanks{P. Vijay Kumar is also an Adjunct Research Professor at the University of Southern California.  This research is supported in part by the National Science Foundation under Grant 1421848 and in part by an India-Israel UGC-ISF joint research program grant.  The work of S. B. Balaji was supported under a TCS research-scholarship program.}
	}
	\maketitle
	
	\begin{abstract}
An erasure code is said to be a code with sequential recovery with parameters $r$ and $t$, if for any $s \leq t$ erased code symbols, there is an $s$-step recovery process in which at each step we recover exactly one erased code symbol by contacting at most $r$ other code symbols.  In earlier work by the same authors, presented at ISIT 2017, we had given a construction for
binary codes with sequential recovery from $t$ erasures, with locality parameter $r$, which were optimal in terms of code rate for given $r,t$, but where the block length was large, on the order of  $r^{c^t}$, for some constant $c >1$.  In the present paper, we present an alternative construction of a rate-optimal code for any value of $t$ and any $r\geq3$, where the block length is significantly smaller, on the order of $r^{\frac{5t}{4}+\frac{7}{4}}$ (in some instances of order $r^{\frac{3t}{2}+2}$).  Our construction is based on the construction of certain kind of tree-like graphs with girth $t+1$. We construct these graphs and hence the codes recursively.
\end{abstract}
	\section{Introduction}
	An $[n,k]$ code $\mathcal{C}$ is said to have locality $r$ if each of the $n$ code symbols of $\mathcal{C}$ can be recovered by contacting at most $r$ other code symbols. Equivalently, there exist $n$ codewords ${\uh_1, \cdots, \uh_n}$, not necessarily distinct, in the dual code $\mathcal{C}^\perp$, such that $i \in \text{supp}(\uh_i)$ and $|\text{supp}(\uh_i)| \leq r+1$ for $1 \leq i \leq n$ where $\text{supp}(\uh_i)$ denotes the support of the codeword $\uh_i$.
	\paragraph{Codes with Sequential Recovery}
	An $[n,k]$ code \calc\ over a field $\mathbb{F}_q$ is defined as a code with sequential recovery \cite{BalPraKum} from $t$ erasures and with locality-parameter $r$, if for any set of $s \leq t$ erased symbols $\{c_{\sigma_1},...,c_{\sigma_s} \}$, there exists a codeword $\uh$ in the dual code \calcp\ of Hamming weight $\leq r+1$, such that $|\text{supp}(\uh) \cap \{\sigma_1,...,\sigma_s \}| = 1$.  We will formally refer to this class of codes as $(n,k,r,t)_{\text{seq}}$ codes. When the parameters $(n,k,r,t)$ are clear from the context, we will simply refer to a code in this class as a code with sequential recovery.
	\subsection{Background}
	In \cite{GopHuaSimYek}, the authors introduced the concept of codes with locality (see also \cite{PapDim,OggDat}), where a symbol is recovered by accessing a subset of $r$ other code symbols. The value of $r$ is typically much smaller than dimension of the code, making the repair process more efficient when compared with MDS codes. The focus of \cite{GopHuaSimYek} was local recovery from single erasure.
	
		There are several approaches to local recovery from multiple erasures (For details please see \cite{BalajiKK17ISIT} and references therein.). Among the class of codes described in \cite{BalajiKK17ISIT} for local recovery from multiple erasures, sequential-recovery codes is the largest class of codes which contains the rest of the class of codes described in \cite{BalajiKK17ISIT}.  For this reason, codes with sequential recovery can potentially achieve higher rate and have larger minimum distance.
		
	The sequential approach to recovery from erasures was introduced by authors in \cite{PraLalKum} and is one of several  approaches to locally recover from multiple erasures. Codes employing this approach have been shown to be better in terms of rate and minimum distance (see \cite{PraLalKum,RawMazVis,SonYue_3_Erasure,SonYue_Binary_local_repair,BalPraKum,balaji2016binary,BalajiKK16,BalajiKK17ISIT}). 
	Local recovery for two erasure case is considered in \cite{PraLalKum} where a tight rate bound for two erasure case and an optimal construction is provided. Codes with sequential recovery from three erasures can be found discussed in \cite{SonYue_3_Erasure,BalPraKum,song2016sequential}. A bound on rate of an $(n,k,r,3)_{seq}$ code was derived in \cite{song2016sequential}.   A rate bound for $t=4$ appears in \cite{balaji2016binary}. The rate problem was completely solved by the authors in \cite{BalajiKK16,BalajiKK17ISIT} where they derived an upper bound on rate and gave a binary construction achieving the rate upper bound for all the parameters $r \geq 3$ and $t$. The downside of the rate-optimal construction in \cite{BalajiKK16,BalajiKK17ISIT} is that it has large block length $O(r^{c^t})$. The main contribution of this paper is to provide a binary construction of rate optimal code with sequential recovery with much smaller block length for all parameters $r \geq 3$ and $t$.


%

%
	\subsection{Contributions of the Paper} 
      \ben
        \item We will first present a method to construct a $d$-regular graph of girth $\geq g+1$ (girth is the length of the smallest cycle in a graph) from a $d$-regular graph of girth $g$ which preserves the structure of the original graph.  Subsequently, we give a method to construct a $d$-regular graph of girth $g$ from a $d$-regular graph of arbitrary girth which preserves the structure of the original graph.
        \item We then use this to construct rate-optimal codes with sequential recovery for parameters $r$ and $t$ from rate-optimal codes with sequential recovery for parameters $r$ and $t'$ (with $t>t', \ (t-t') \text{ even}$) using graph-theoretic methods that are introduced here. All codes constructed in this paper are binary codes. 
        \item  We also give a unified viewpoint of construction by unifying the construction for $t$ even and $t$ odd case whereas in \cite{BalajiKK16}, the $t$ even and $t$ odd cases were treated differently. 
        \item Apart from that our construction reduces the block length to $O(r^{\frac{5t}{4}+\frac{7}{4}})$ (in some instances block length is $O(r^{\frac{3t}{2}+2})$) which is considerably less compared to the construction in \cite{BalajiKK16} which is of block length $O(r^{c^{t}})$ (for some $c>1$). Although our block length is still $O(r^{\frac{5t}{4}+\frac{7}{4}})$, much improvement cannot be expected in block length as there is a lower bound on block length of $ O(r^{\frac{t+1}{2}})$.
    \een
    
	\section{A Method to Construct a $d$-Regular Graph of Girth $\geq g+1$ from a $d$-Regular Graph of Girth $g$ preserving the structure:} \label{graph_girth}
	\subsection{Case: $g$ is Odd}
	Let $G=(V,E)$ be a $d$-regular graph of girth $g$ with vertex set $V=\{v_1,\hdots,v_m\}$ and edge set $E$.
	
	{\em Steps to convert a $d$-regular graph of girth $g$ to a $d$-regular graph of girth $\geq g+1$:}
	\ben
	\item Replace each vertex $v_i \in V$ in $G$ with two vertices $v_{(i,0)},v_{(i,1)}$ to form a graph $G_1$.
	\item For every edge $(v_i,v_j) \in E$ in $G$, form two edges $(v_{(i,0)},v_{(j,1)})$, $(v_{(i,1)},v_{(j,0)})$ in $G_1$.
	\een
	{\em $G_1$ is a $d$-regular graph with girth $\geq g+1$:}
	\ben
	\item {\em $G_1$ is a $d$-regular graph: } Every vertex of graph $G_1$ has degree $d$ because if the neighbours of $v_i$ in $G$ are $\{v_{i_1},\hdots,v_{i_d} \}$ then the neighbours of the vertex $v_{(i,0)}$ in $G_1$ is $\{v_{(i_1,1)},\hdots,v_{(i_d,1)} \}$. Hence $\text{deg}(v_{(i,0)})=d$. Similarly the neighbours of the vertex $v_{(i,1)}$ in $G_1$ is $\{v_{(i_1,0)},\hdots,v_{(i_d,0)} \}$. Hence $\text{deg}(v_{(i,1)})=d$.
	\item {\em $G_1$ has girth $\geq g+1$: }  Suppose there is a cycle $C$ of length $g$ in $G_1$. Then there should be a vertex $v_{(i_1,0)}$ in $C$ for some $i_1$. Let the other vertices in the cycle in sequence be $v_{(i_2,1)},v_{(i_3,0)},...,v_{(i_g,1)}$. Note that the `last' vertex should be $v_{(i_g,1)}$ as there is an edge from $v_{(i_g,1)}$ to $v_{(i_1,0)}$. However this means $g$ should be even, which is not the case. Hence cycle of length $g$ is not possible in $G_1$. Further since any cycle of length $<g$ in $G_1$ must correspond to a cycle of length $<g$ in $G$, cycles of length $<g$ are not present in $G_1$. Hence girth of $G_1$ is $\geq g+1$. In particular $G_1$ has no odd cycles.
	\een
	
	\subsection{Case: $g$ is Even}
	
		Let $G=(V,E)$ be a $d$-regular graph of girth $g$ with vertex set $V=\{v_1,\hdots,v_m\}$ and edge set $E$. Let $H$ be a group (for example $H$ could be a vector space) such that $|H|>(d-1)^{\frac{g}{2}}$.
		
		{\em Steps to convert a $d$-regular graph of girth $g$ to a $d$-regular graph of girth $\geq g+1$:}
		\ben
		\item Replace each vertex $v_i \in V$ in $G$ with the set $\{v_{(i,h)} : h \in H \}$ of vertices to form a graph $G_1$.
		\item For every edge $(v_i,v_j) \in E$ in $G$, form $|H|$ edges $(v_{(i,h)},v_{(j,hh_{(i,j)})})$, $\forall h \in H$  in $G_1$ for a chosen $h_{(i,j)}$.  We here observe that $h_{(i,j)} = h_{(j,i)}^{-1}$.
				\item {\em $G_1$ is a $d$-regular graph: } Every vertex of graph $G_1$ has degree $d$ because if the neighbours of $v_i$ in $G$ are $\{v_{i_1},\hdots,v_{i_d} \}$ then the neighbours of the vertex $v_{(i,h)}$ in $G_1$ is $\{v_{(i_1,hh_{(i,i_1)})},\hdots,v_{(i_d,hh_{(i,i_d)})} \}$. Hence $\text{deg}(v_{(i,h)})=d$.
		\item In the following we will see on how to choose $h_{(i,j)}$ such that the graph $G_1$ has girth $\geq g+1$.
		\een
		{\em Choosing the set $\{h_{(i,j)}\}$ such that $G_1$ is a $d$-regular graph with girth $\geq g+1$:}
		\ben
		\item {\em $G_1$ has girth $\geq g+1$: } Let wolog $v_{(i_1,h)},v_{(i_2,hw_1)}$ $,\hdots,v_{(i_g,hw_{g-1})}$ be a set of $g$ vertices such that $(v_{(i_J,hw_{J-1})},v_{(i_{J+1},hw_J)})$  is an edge in $G_1$ $\forall 1 \leq J \leq g-1$ and  $(v_{(i_g,h w_{g-1})},v_{(i_{1},h)})$ is an edge in $G_1$ where  $w_J= \prod_{j=1}^{J}h_{(i_j,i_{j+1})}$,$\forall 1 \leq J \leq g-1$. Hence the vertices $v_{(i_1,h)},v_{(i_2,hw_1)},\hdots,v_{(i_g,h w_{g-1})}$ form a cycle of length $g$. For this cycle to be in $G_1$, we must have $ w_{g-1} h_{(i_g,i_1)}=e$ where $e$ is the identity element in $H$. Hence if the element $h_{(i_g,i_1)}$ is different from $w_{g-1}^{-1}$, then the above cycle would not occur in $G_1$. Hence for every cycle in $G$ of length $g$ involving the edge $(v_i,v_j)$, we must avoid $h_{(i,j)}$ being equal to precisely one element from $H$ to avoid the cycle of $G$ from getting carried over to $G_1$. By the construction of $G_1$ from $G$, it is clear that every cycle of $G_1$ of length $g$ corresponds to a cycle of $G$ of length $g$ i.e., if $v_{(i_1,h)},v_{(i_2,hw_1)}$ $,\hdots,v_{(i_g,h w_{g-1})}$ form a cycle as above then $v_{i_1},v_{i_2}$ $,\hdots,v_{i_g}$ must form a cycle in $G$. Hence it is enough to avoid the cycles of $G$ of length $g$ from getting carried over to $G_1$ by choosing $\{h_{(i,j)}\}$ such that it avoids all cycles of length $g$. 
		\item We have already seen that to avoid cycle of length $g$ in $G_1$, $h_{(i,j)}$ must not equal precisely one element from $H$ for every cycle of length $g$ involving the edge $(v_i,v_j)$ in $G$. If we prove that the number of cycles of length $g$ involving a given edge in $G$ is less than $|H|$, then there definitely exists a choice of $\{h_{(i,j)}\}$ such that it avoids all cycles of length $g$.
		\item {\em Counting the maximum number of cycles of length $g$ involving a given edge $(v_i,v_j)$ in $G$:} Let us take a vertex $v_i \in V$ in $G$ and take all its $d$ neighbours, let these neighbours form a set $N_1$. Let us take all the neighbours of vertices in $N_1$ apart from $v_i$ and form the set $N_2$ with these vertices. Repeat the argument: at step $i$, take all neighbours of the vertices in the set $N_{i-1}$ apart from the vertices in $N_{i-2}$ and form the set $N_i$ with these vertices. Since the girth of $G$ is $g$, the sets $N_1,\hdots,N_{\frac{g}{2}-1}$ are all pairwise disjoint and $|N_i|=d(d-1)^{i-1}$. Now take a neighbour of $v_i$, (say) $v_j$ and do the above procedure (leaving out $v_i$ from the neighbour set of $v_j$ in the first step) and form sets $M_1,\hdots,M_{\frac{g}{2}-1}$. Note that $M_i \subseteq N_{i+1}$, $\forall 1 \leq i \leq \frac{g}{2}-2$ and $M_{\frac{g}{2}-1}$ is disjoint with any set among $N_1,\hdots,N_{\frac{g}{2}-1}$ as otherwise we would have a cycle of length $<g$. This is depicted in Fig~\ref{fig:Cycles_of_length_g}. Now each vertex in $M_{\frac{g}{2}-1}$ can possibly connect via edges to $d-1$ distinct vertices in the set $N_{\frac{g}{2}-1} - M_{\frac{g}{2}-2}$ where each such edge will be present in unique cycle of length $g$ involving the edge $(v_i,v_j)$. Also any cycle of length $g$ involving the edge $(v_i,v_j)$ must involve an edge between a vertex in  $M_{\frac{g}{2}-1}$ and a vertex in $N_{\frac{g}{2}-1} - M_{\frac{g}{2}-2}$. Fig~\ref{fig:Cycles_of_length_g} depicts this. Hence the maximum number of cycles of length $g$ containing the edge $(v_i,v_j)$ which are possible is $ \leq (d-1)|M_{\frac{g}{2}-1}|=(d-1)^{\frac{g}{2}}$.
		\item Since $|H|>(d-1)^{\frac{g}{2}}$ ($|H|>$ maximum number of cycles of length $g$ involving an arbitrary edge $(v_i,v_j)$ in $G$), we can choose $\{h_{(i,j)}\}$ to avoid all the cycles of length $g$ in $G_1$. Hence $G_1$ can be constructed with girth $\geq g+1$.
		\een
		Since the graphs are constructed from $G$ by replacing a vertex with multiple vertices and defining edges closely following the edges of $G$, the graph $G_1$ preserves some of the structure of $G$. What exactly we mean by preserving the structure will be clear when we use this construction to construct codes with sequential recovery with optimal rate.
		
	\section{A Method to Construct a $d$-Regular Graph of Girth $\geq g$ from a $d$-Regular Graph of Arbitrary Girth preserving the structure:} \label{graph_girth_cayley}
			Let $G=(V,E)$ be a $d$-regular graph of arbitrary girth with vertex set $V=\{v_1,\hdots,v_{|V|}\}$ and edge set $E$. Let $H$ be a group. We will choose $H$ in the procedure described below.
			
	\ben
	  \item We define the graph in the same way as last section:
	  	\ben
	  	\item Replace each vertex $v_i \in V$ in $G$ with the set $\{v_{(i,h)} : h \in H \}$ of vertices to form a graph $G_1$.
	  	\item For every edge $(v_i,v_j) \in E$ in $G$, form $|H|$ edges $(v_{(i,h)},v_{(j,hh_{(i,j)})})$, $\forall h \in H$  in $G_1$ for a chosen $h_{(i,j)}$.  We here observe that $h_{(i,j)} = h_{(j,i)}^{-1}$.
	  	\item In the following we will see on how to choose $h_{(i,j)}$ such that the graph $G_1$ has girth $g$.
	  	\een
	  \item Let us colour the edges of the graph $G$ with least number of colours such that no two adjacent edges have the same colour. By Vizing's theorem, we can do such colouring with at most $d+1$ colours. Let the colours used be $\{c_1,\hdots,c_{d+1}\}$.
	  \item Choose group $H$ such that its undirected Cayley graph $\text{Cay}(H,S)$ relative to a symmetric set of elements $S$ has girth $\geq g$. The vertex set of Cayley graph $\text{Cay}(H,S)$ is the group elements $H$ and the edge set is $\{(h,hs): h \in H, s\in S\}$ and $S$ is such that if $s \in S$ then $s^{-1} \in S$. We choose $\text{Cay}(H,S)$ such that $|S| \geq 2(d+1)$.
	  \item A sequence of $m$ elements from $S$ : $s_1,\hdots,s_m$ is said to be a reduced word of length $m$ if $s_i \neq s_{i+1}^{-1}$, $\forall 1 \leq i \leq m-1$. Since the Cayley graph $\text{Cay}(H,S)$ has girth $\geq g$, any reduced word of length $<g$ is not equal to $e$ (identity element of $H$).
	  \item Choose $S_1 \subseteq S$ such that $|S_1| = \lceil |S|/2 \rceil $ and if $s \in S_1$ then $s^{-1} \notin S_1$ (If $s \in S$ is such that $s=s^{-1}$, we allow that $s$ in $S_1$). Such a choice of subset is always possible by uniqueness of inverse.
	  \item Let $L=|S_1|$, $S_1=\{s_1,\hdots,s_L\}$. If the colour of the edge $(v_i,v_j)$ is $c_k$ then we set $h_{(i,j)} = s_k$. Since $L \geq d+1$ such an assignment is possible.
	  \item Let wolog $v_{(i_1,h)},v_{(i_2,hw_1)}$ $,\hdots,v_{(i_m,hw_{m-1})}$ be a set of $m$ vertices such that $(v_{(i_J,hw_{J-1})},v_{(i_{J+1},hw_J)})$  is an edge in $G_1$ $\forall 1 \leq J \leq m-1$ and  $(v_{(i_m,h w_{m-1})},v_{(i_{1},h)})$ is an edge in $G_1$ where  $w_J= \prod_{j=1}^{J}h_{(i_j,i_{j+1})}$,$\forall 1 \leq J \leq m-1$. Hence the vertices $v_{(i_1,h)},v_{(i_2,hw_1)},\hdots,v_{(i_m,h w_{m-1})}$ form a cycle of length $m$. For this cycle to be in $G_1$, we must have $ w_{m-1} h_{(i_m,i_1)}=e$ i.e., $ \prod_{j=1}^{m-1}h_{(i_j,i_{j+1})} \times h_{(i_m,i_1)} = e$. But $\prod_{j=1}^{m-1}h_{(i_j,i_{j+1})} \times h_{(i_m,i_1)}$ is a reduced word of length $m$ (It is reduced because of the edge colouring and choice of $S_1$ such that if $s \in S_1$ then $s^{-1} \notin S_1$.). Hence $m>=g$. Hence girth of $G_1$ $\geq g$.
	  \item Note that $G_1$ just inherits the girth of $\text{Cay}(H,S)$ and has $|G| \times |H|$ vertices.Although we are getting the girth property directly from $\text{Cay}(H,S)$, the main point of this construction is that the graph $G_1$ inherits the structure of $G$ and girth of $\text{Cay}(H,S)$. We will see later that inheriting the structure of $G$ will help in the construction of codes with sequential recovery.
	\een
	Note that the above construction uses a girth $g$ graph (Cayley graph) of high degree to construct a girth $g$ graph of low degree from a graph $G$ of low degree and low girth. This may sound strange as we could have taken the high girth and high degree Cayley graph as our final graph but the Cayley graph as such is not useful for us for constructing rate-optimal codes with sequential recovery. For constructing rate-optimal codes with sequential recovery, we need a tree-like structure in the graph we use to define the code. Hence we want to construct a high girth graph while preserving the structure of an initial graph $G$. Hence our construction mentioned in this section is important for constructing high girth graph while preserving the structure of an initial graph $G$. This will become more clear when we construct our codes. It will be interesting to study the various graph properties $G_1$ inherits from $G$ apart from the properties we need for our code construction. For example, the independence number of $G_1$ $\geq $ independence number of $G$ $\times$ cardinality of $H$. We can also construct $(d_v,d_c)$-regular LDPC code of high girth using the method developed in this section where $G$ will be the Tanner graph of a $(d_v,d_c)$-regular LDPC code of low girth (Although we said $G$ is a regular graph in our method, $G$ can actually be any graph with $d$ playing the role of maximum degree.).
    
	Note that if we want to go from a graph of girth $g$ to a graph of girth $g+1$ while inheriting the structure of the original graph then the method described in the last section will give lesser number of vertices as the final graph will have $ |G| \times |H|$ vertices where $|H| \approx (d-1)^{\frac{g}{2}}$. The method described in this section will also have $ |G| \times |H|$ vertices but $|H|$ is equal to smallest number of vertices possible in a Cayley graph of degree at least $2(d+1)$ and girth $g$ which is at least $(2d+1)^{\frac{g}{2}}$ .The method described in this section is more efficient in terms of number of vertices if we want to go from very small girth to girth $g$ while preserving the structure of original graph. For example if we use the Cayley graph construction given in \cite{Lubotzky1988} and if $p,q$ are primes such that  $p+1 \geq 2(d+1)$ and $q \geq 4p^{\frac{g}{4}}$ and $p,q = 1 \text{ mod } 4$ and $p$ is not a quadratic residue of $q$ then $\text{Cay}(PGL(2,\mathbb{Z}_q),S)$ will have girth $\geq g$ with $q(q^2-1)$ vertices where $S$ is a set of $p+1$ generators of $PGL(2,\mathbb{Z}_q)$ which are image of a homomorphism between Quaternions over integers and $PGL(2,\mathbb{Z}_q)$ based on solutions of $a_0^2+a_1^2+a_2^2+a_3^2 = p$ where $a_i \in \mathbb{Z}$. Hence if we use $PGL(2,\mathbb{Z}_q)$, then we can construct a graph of degree $d$ and girth $\geq g$ from a graph $G$ of arbitrary girth and degree $d$ with number of vertices of the final graph being $\approx 64(2d+1)^{\frac{3g}{4}} \times |G|$ which is less then repeated application of the method described in last section. For Cayley graph with less vertices with girth $\geq g$ for a larger range of parameters (i.e., without the need to search for primes $p,q$ as mentioned before), please refer to \cite{X_Dahan,Mor,DavSarVal}.
	
	We described the above method because of its simplicity. There is a method to construct a graph of girth $\geq g$ with degree $d$ from $G$ by using any graph of degree $d+1$ and girth $\geq g$ while preserving the structure of $G$. This will yield even lesser number of vertices than the above method. We only briefly outline the method.

\subsection{Outline of a Generic Method to Construct a $d$-Regular Graph of Girth $\geq g$ from a $d$-Regular Graph of Arbitrary Girth Preserving the Structure } \label{girth_arb_graph}

	\ben
	\item Construction of $d$ regular graph of girth $\geq g$ from a $d$ regular graph of arbitrary girth preserving the structure can be done using any graph of degree $d+1$ and girth $\geq g$ rather than using the Cayley graph of degree $\geq 2(d+1)$ which was used in the previous construction.
	\item Let $G=(V,E)$ be a $d$-regular graph with arbitrary girth. Take a graph $G'=(V',E')$ of degree $d+1$ and girth $\geq g$. Construct $d+1$ matchings $\{M_i: 1 \leq i \leq d+1 \}$ of $G'$ such that $E'=\cup_i M_i$ and $M_i \cap M_j =\emptyset$ and the edges in $M_i$ are a set of edges such that each vertex of $G'$ appears exactly once in exactly one edge in $M_i$.
	\item If we cannot construct such matchings then there is a standard technique to convert $G'$ into a bipartite graph with twice the number of vertices and degree $d+1$ and girth $\geq g$. We will replace $G'$ with this new bi-partite graph. By repeated application of Hall's marriage theorem, we know the existence of matchings $\{ M_i: 1\leq i \leq d+1\}$ in this new bipartite graph $G'$.
     \item Let us colour the edges of the graph $G$ with least number of colours such that no two adjacent edges have the same colour. By Vizing's theorem, we can do such colouring with at most $d+1$ colours. Let the colours used be $\{c_1,\hdots,c_{d+1}\}$.
	\item Now replace each vertex $v_i \in V$ by $|V'|$ vertices $\{v_{(i,w)} : w \in V'\}$. Call the new graph $G_1$. The edges in $G_1$ are defined as follows:
	\ben
	 \item For each edge $(v_i,v_j) \in E$, (let wolog the colour of the edge be $c_k$) form edges $\{(v_{(i,w_a)},v_{(j,w_b)}),(v_{(i,w_b)},v_{(j,w_a)}) : (w_a,w_b) \in M_k \}$.
	\een
	\item It is clear that graph $G_1$ has degree $d$. The fact that it has girth $\geq g$ can be seen as follows.
	\ben
	 \item Wolog let the vertices: $v_{(i_1,w_1)},v_{(i_2,w_2)},\hdots,$ $v_{(i_{m-1},w_{m-1})},v_{(i_m,w_m)}$ form a cycle of length $m$. Now $w_1,w_2,\hdots,w_{m-1},w_m$ must form a cycle in $G'$ because the construction is such that the edge $(w_j,w_{j+1})$ is not equal to edge $(w_{j+1},w_{j+2})$ $,\forall 1 \leq j \leq m-2$ due to colouring and assignment of distinct matching to distinct colouring and the fact if $(v_{(i,w_a)},v_{(j,w_b)})$ is an edge then $(v_{(i,w_b)},v_{(j,w_a)})$ is also an edge, as if  $(w_j,w_{j+1}) = (w_{j+1},w_{j+2})$ and if colour of $(v_{i_j},v_{i_{j+1}})$ is $c_{k_1}$ and colour of $(v_{i_{j+1}},v_{i_{j+2}})$ is $c_{k_2}$ (by the construction it is clear that $(v_{i_j},v_{i_{j+1}}) \neq (v_{i_{j+1}},v_{i_{j+2}})$) then $(w_{j},w_{j+1}) \in M_{k_1}$ $(w_{j+1},w_{j+2}) \in M_{k_2}$ which is a contradiction as $k_1 \neq k_2$ and the matchings are disjoint. Since girth of $G'$ is $\geq g$, $m \geq g$. Hence $G_1$ has girth $\geq g$.
	\een
	\een

	\begin{figure}[ht]
		\centering
		\includegraphics[width=6cm,height=4cm]{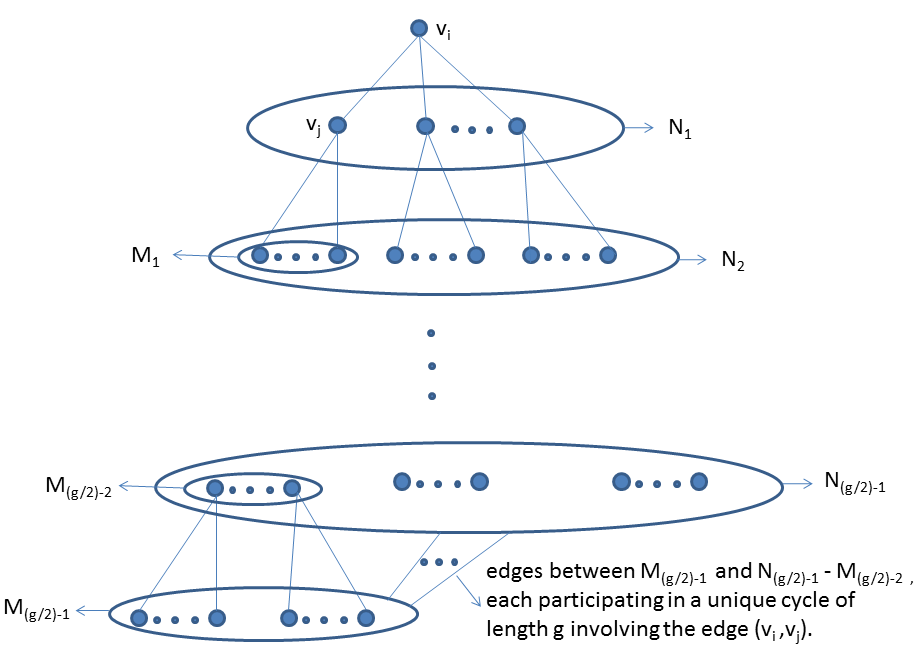}
		\caption{Figure showing all possible cycles of length $g$ involving the edge $(v_i,v_j)$.}
		\label{fig:Cycles_of_length_g}
	\end{figure}
	
	\section{Construction of Codes with Sequential Recovery for any $r$ and $t$ with Optimal Rate:}
	We now describe our construction of code with sequential recovery for a given $r$ and $t$ with optimal rate over $\mathbb{F}_2$. Before we describe the actual code, we describe a graph $G_{s-1}$ where $s=\frac{t}{2}$ for $t$ even and $s=\frac{t-1}{2}$ for $t$ odd. Our construction of the code will be based on this graph $G_{s-1}$.
	From \cite{BalajiKK16}, it can be seen that a code with sequential recovery for a given $r,t$ with optimal rate over binary field need to have the structure of the code we describe in the following. In \cite{BalajiKK16}, we gave a completely different construction for $t$ even and $t$ odd. But here in this paper, we give a unified viewpoint and unified construction for both $t$ even and $t$ odd. Apart from giving a unified construction, the block length of our construction is small compared to the rate-optimal construction given in \cite{BalajiKK16}.
	
	\subsection{Description of Graph $G_{s-1}$} \label{Gs}
	The graph we are going to describe builds upon a base graph $G_0$.
	
	\ben
	\item Take a graph $G_0$ (will be defined later) with the set of vertices partitioned into two sets $U_0,L_0$.
	\item {\em Step 1: }Now partition the set of vertices in $U_0$ into sets of size exactly $r$. Let the partition be $P^0_1,\hdots,P^0_{\frac{|U_0|}{r}}$. Now add a vertex $u^0_i$ to $G_0$ such that $u^0_i$ is connected via edges to vertices in $P^0_i$, $\forall 1 \leq i \leq \frac{|U_0|}{r}$. Let the resulting graph be called $G_1$ and set $U_1=\{u^0_1,\hdots,u^0_{\frac{|U_0|}{r}}\}$.
	\item {\em Step i: } Let the graph constructed at step $i-1$ be $G_{i-1}$ with a subset of vertices defined in the step $i-1$ called $U_{i-1}$. Now partition the set of vertices in $U_{i-1}$ into sets of size exactly $r$. Let the partition be $P^{i-1}_1,\hdots,P^{i-1}_{\frac{|U_{i-1}|}{r}}$. Now add a vertex $u^{i-1}_j$ to $G_{i-1}$ such that $u^{i-1}_j$ is connected via edges to vertices in $P^{i-1}_j$, $\forall 1 \leq j \leq \frac{|U_{i-1}|}{r}$. Let the resulting graph be called $G_i$ and set $U_i=\{u^{i-1}_1,\hdots,u^{i-1}_{\frac{|U_{i-1}|}{r}}\}$.
	\item Do step $i$ for $1 \leq i \leq s-1$ and construct the graph $G_{s-1}$ with a subset of vertices defined in step $s-1$ called $U_{s-1}$.
	\item {\em Definition of $G_0$:} \ben
		\item For $t$ even, $G_0$ is a $r$-regular graph with $U_0=V(G_0)$, $L_0 = \emptyset$ where $V(G_0)$ is the vertex set of $G_0$.
		\item For $t$ odd, $G_0$ is a bipartite graph with the two sets of vertices corresponding to $U_0$ and $L_0$ (vertices in $U_0$ has no edges among them and vertices in $L_0$ has no edges among them). All the vertices in $U_0$ has degree $r$ and all the vertices in $L_0$ has degree $r+1$.
		\item $G_0$ is chosen such that it satisfies the above definition and as well as such that it ensures that the graph $G_{s-1}$ has girth $\geq t+1$. To ensure this both $G_0$ and the sets in the partition $P_j^i$ must be chosen carefully. Note that for constructing $G_{s-1}$, the number of nodes in $U_0$ must be a multiple of $r^{s-1}$.
	\een
	\een     
	\subsection{Description of the Code with Sequential Recovery for a given $r$ and $t$ with Optimal Rate from the Graph $G_{s-1}$}\label{code}
	\ben
	 \item Let $|U_{s-1}|=a_0$. Let $U_{s-1}=\{u_1^{s-2},\hdots,u_{a_0}^{s-2} \}$. Add new vertices $w_1,\hdots,w_{a_0}$ to the graph $G_{s-1}$ with $w_i$ connected via an edge to $u_i^{s-2}$, $\forall 1 \leq i \leq a_0$. Let the resulting graph be $G_{s-1}^1$.
	 \item Now let each edge in $G_{s-1}^1$ represent a unique code symbol of our code and let each vertex except vertices in the set $\{w_1,\hdots,w_{a_0}\}$ represent a parity check of the code symbols corresponding to the edges incident on it.
	 \item Note that the edge between $w_i$ and $u_i^{s-2}$ does represent a unique code symbol of our code but the vertices $w_i$ do not represent a parity check. The vertices $w_i$ are introduced as dummy nodes to introduce the code symbols represented by the edges between $w_i$ and $u_i^{s-2}$.
	 \item Hence the code is defined by the code symbols corresponding to edges in $G^1_{s-1}$ and parity checks corresponding to vertices.
	 \item {\em Rate of the constructed code:}
	 \ben
	 \item {\em Rate of our code for $t$ even:}
	   \bean
	     n-k & \leq & \text{$e_1$ = number of nodes in $G^1_{s-1}$} \\
	     	& & \text{except $\{w_i\}$ } \\
	     &=& \sum_{j=0}^{s-1} \frac{|G_0|}{r^j}
	     \eean
	     \bean
	     n&=& \text{ number of edges in $G^1_{s-1}$} \\
	     & =& \frac{e_1 \times (r+1)}{2}+\frac{|G_0|}{2r^{s-1}}
	     \eean
	     \bean
	     \text{ Hence: rate } = \frac{k}{n} &\geq& \frac{r^s}{r^s+2\sum_{j=0}^{s-1}r^i}.
	   \eean
	   Hence rate of the code described meets the rate upper bound given in \cite{BalajiKK16}. Hence the code we described is rate-optimal.
	 \item {\em Rate of our code for $t$ odd:}
	 \bean
	 n-k & \leq & \text{$e_1$ = number of nodes in $G^1_{s-1}$}\\
	       & & \text{except $\{w_i\}$ } \\
	 &=& \sum_{j=0}^{s-1} \frac{|U_0|}{r^j} + \frac{r|U_0|}{r+1} \\
	 \eean
	 \bean
	 n&=& \text{ number of edges in $G^1_{s-1}$} \\
	 & =& \frac{e_1 \times (r+1)}{2}+\frac{|U_0|}{2r^{s-1}}
	 \eean
	 \bean
	 \text{Hence: rate } = \frac{k}{n} &\geq& \frac{r^{s+1}}{r^{s+1}+2\sum_{j=1}^{s}r^i+1}.
	 \eean
	 Hence rate of the code described meets the rate upper bound given in \cite{BalajiKK16} (Note that in \cite{BalajiKK16}, $s=\frac{t+1}{2}$ but here $s=\frac{t-1}{2}$). Hence the code we described is rate-optimal.
	 \een
	\een
	The fact that the code described can correct $t$ erasures sequentially follows from the argument in \cite{BalajiKK16}. The sequential recovery from $t$ erasures is possible due to the fact that $G_{s-1}$ has girth of $t+1$ and degree one code symbols are well separated in the Tanner graph. This is due to the tree-like structure of $G^1_{s-1}$ as the edges between $w_i$ and $u^{s-2}_i$ $\forall 1 \leq i \leq a_0$ are precisely the code symbols of degree one in Tanner graph. To go from one degree-one code symbol to another degree-one code symbol, one needs to go down the tree to last depth and come up the tree (The part of $G^1_{s-1}$ defined by partitions $P_j^i$ is the tree-like structure of the graph $G^1_{s-1}$.) and hence well separated. These degree one code symbols are the reason for increase in rate leading to rate-optimality. We skip detailed arguments here as it would be a repetition of arguments in \cite{BalajiKK16}.
	
	\subsection{Construction of Code with Sequential Recovery with parameters $r$ and $t$ from a Code with Sequential Recovery with parameters $r$ and $t'$ for any $t'< t$ such that $t-t'$ is even}
	
	From the previous subsection, it is clear that to construct a code with sequential recovery with parameters $r$ and $t$ from a code with sequential recovery with parameters $r$ and $t'$ ($t-t'$ is even), it is enough to construct $G_{s-1}$ with girth $\geq t+1$ from $G_{s'-1}$ with girth $\geq t'+1$ where $G_{s'-1}$ has the tree-like structure and constructed as described in Section \ref{Gs} starting from a base graph $G_0$  where $s'=\frac{t'}{2}$ for $t'$ even and $s'=\frac{t'-1}{2}$ for $t'$ odd.
	
	{\em Construction of $G_{s-1}$ of girth $\geq t+1$ from $G_{s'-1}$ of girth $\geq t'+1$:}
	\ben
	\item {\em Step 1: } Take the set of vertices in $U_{s'-1}$ in $G_{s'-1}$ and partition it into sets of size exactly $r$. Let the partition be $P^{s'-1}_1,..,P^{s'-1}_{\frac{|U_{s'-1}|}{r}}$. The choice of partition is arbitrary. Note that $|U_{s'-1}|$ must be a multiple of $r$ for forming the partition. If not, we replicate $G_{s'-1}$ and take union of copies of $G_{s'-1}$ until $|U_{s'-1}|$ is a multiple of $r$. Now add new vertex $u^{s'-1}_i$ and connect it via edges with vertices in $P^{s'-1}_i$, $\forall 1 \leq i \leq \frac{|U_{s'-1}|}{r}$. Let the resulting graph be called $G^{temp}_{s'}$ and set $U_{s'} = \{u^{s'-1}_1,\hdots,u^{s'-1}_{\frac{|U_{s'-1}|}{r}}\}$.
	\item Repeat the previous step (step 1) with $G^{temp}_{s'}$ as the starting graph to get $G^{temp}_{s'+1}$. Keep doing this i.e., repeat (step 1) until we get $G^{temp}_{s-1}$.
	\item Now $G^{temp}_{s-1}$ has the necessary structure for constructing rate optimal code for parameters $r$ and $t$ except that girth of $G^{temp}_{s-1}$ is guaranteed to be only $t'+1$.
	\item Now apply the method described in Section \ref{graph_girth_cayley} to the graph $G^{temp}_{s-1}$ to construct a new graph $G_{s-1}$ of girth $\geq t+1$.
	\item Since the method described in Section \ref{graph_girth_cayley}, just replaces a vertex with multiple vertices and preserves the structure of neighbours from the original graph $G^{temp}_{s-1}$ in a certain way (The method just replaces each vertices with multiple vertices and if there is an edge between two vertices then it replaced by a matching between the vertices which replaced the two vertices. Hence the method preserves the tree-like structure of $G^{temp}_{s-1}$.), the graph $G_{s-1}$ also has tree-like structure (inherited from graph $G^{temp}_{s-1}$) and can be constructed starting from the base graph $G^{t+1}_0$ as described in this Section \ref{Gs} where $G^{t+1}_0$ is just the graph $G_0$ with the method described in Section \ref{graph_girth_cayley} applied on it to have girth $\geq t+1$. Note that the method described in Section \ref{graph_girth_cayley} also applies to graph with irregular degrees. Hence it can be applied to the graphs $G^{temp}_{s-1}$ and $G_0$.
	\item Hence we have constructed the graph $G_{s-1}$ with required properties. Now defining the code on this graph $G_{s-1}$ as described in Section \ref{code} will give a rate-optimal code with sequential recovery with parameters $r$ and $t$.
	\item The number of vertices in $G_{s-1}$ is $|H| \times |G^{temp}_{s-1}|$  where $H$ is group used while applying the method of Section \ref{graph_girth_cayley} on $G^{temp}_{s-1}$. Note that $|H|\approx O((2r+3)^{\frac{3(t+1)}{4}})$ (as explained in Section \ref{graph_girth_cayley} for some family of parameters $r$ and $t$). Hence the number of vertices in $G_{s-1}$ is $\approx O((2r+3)^{\frac{3(t+1)}{4}}) \times |G^{temp}_{s-1}|$ and $|G^{temp}_{s-1}| \approx O(r^{s})$ (if we start with a graph $G_{s'-1}$ with very small girth $t'+1$ i.e., $t'$ is very small and construct $G_{s'-1}$ properly.) and hence the resulting code will have block length $O((2r+3)^{\frac{5t}{4}+\frac{7}{4}})$.  Note that the construction described in \cite{BalajiKK16} needs approximately $O(r^{c^{t}})$ vertices  for some constant $c>1$ and hence very large block length. Our construction described in this paper reduces this considerably by bringing down the exponent of $r$ from $c^{t}$ to $\frac{5t}{4}+\frac{7}{4}$. The block length of $O((2r+3)^{\frac{5t}{4}+\frac{7}{4}})$ can be further reduced to $O((r+2)^{\frac{5t}{4}+\frac{7}{4}})$, if we use a graph $G'$ of girth $\geq t+1$ and degree $r+2$ with $O(r^{\frac{3(t+1)}{4}})$ vertices (See \cite{LazUstWol} for example) and apply  the method described in Section \ref{girth_arb_graph} to construct $G_{s-1}$ from $G^{temp}_{s-1}$. In worst case our block length is $O(r^{\frac{3t}{2}+2})$, since there is a construction of graph of degree $r+1$ and girth $\geq t+1$ (\cite{X_Dahan}) with vertices of order $O((r+1)^{t})$.
	\item If $t=t'+2$ then we apply the method described in Section \ref{graph_girth} rather than the method in  Section \ref{graph_girth_cayley} to construct $G_{s-1}$ from $G^{temp}_{s-1}$ which will have lesser number of vertices and hence lesser block length than that described in previous point.
	\item Note that one possible way to construct $G_{s-1}$ of girth $t+1$ is to take a graph $G_2$ of girth $\geq t+1$ and degree $r+1$ and remove several carefully chosen vertices so that a certain $a_0$ vertices become of degree $r$ and expand the neighbourhood structure of these $a_0$ vertices like a tree up to depth $0$ (where $a_0$ vertices are at depth $s-1$). The difficulty with this approach is in choosing $a_0$ vertices such that we can expand up to required depth without repetition of vertices and the difficulty also lies in the fact that all the vertices in $G_2$ (apart from the once we removed at beginning) must be present in this expansion i.e., no extra vertices must be present in $G_2$ after expanding $a_0$ vertices up to depth $0$ (we cannot remove the extra vertices as this would change the property of base graph and hence would not yield a rate-optimal code). Although it is difficult to follow this approach mentioned in this point, there is a small class of graphs for which the approach works. These graphs are called Moore graphs. But Moore graphs form a very small class of graphs with much restricted parameters. For these reasons, we do not adopt this approach and follow the method we described in this paper.
	\item From the Moore bound (bound based on the idea of Moore graph) (See \cite{DynCageSur}), we can see that block length has to be greater than or equal to $\approx r^{\frac{t+1}{2}}$. Hence we have block length reasonably close to optimal.
	\een
\bibliographystyle{IEEEtran}
\bibliography{bib_file}	
\end{NoHyper}
\end{document}